%% file: paper.tex
\documentstyle[prb,aps,amsfonts,amstex,amssymb,floats,exscale,epsfig]{revtex}

\begin{document}

\makeatletter

\title{Ion-ion correlations: an improved one-component plasma correction}
\author{Marcia C. Barbosa$^1$,  Markus Deserno$^2$ and Christian Holm$^2$}
\address{$^1$ Instituto de F{\'\i}sica, Universidade Federal do Rio 
         Grande do Sul, Caixa Postal 15051, \\ 91501-970 Porto Alegre 
         (RS), Brazil \hspace{1ex} email: {\sf barbosa@if.ufrgs.br}}
\address{$^2$ Max-Planck-Institut f{\"u}r Polymerforschung, Ackermannweg 10, 
         55128 Mainz, Germany \\ email: {\sf deserno@mpip-mainz.mpg.de,
         holm@mpip-mainz.mpg.de}}
\date{22 October 1999}

\makeatother

\maketitle

\begin{abstract}
  Based on a Debye-H{\"u}ckel approach to the one-component plasma we
  propose a new free energy for incorporating ionic correlations into
  Poisson-Boltzmann like theories. Its derivation employs the
  exclusion of the charged background in the vicinity of the central
  ion, thereby yielding a thermodynamically stable free energy
  density, applicable within a local density approximation. This is an
  improvement over the existing Debye-H{\"u}ckel plus hole theory,
  which in this situation suffers from a ``structuring
  catastrophe''. For the simple example of a strongly charged stiff
  rod surrounded by its counterions we demonstrate that the
  Poisson-Boltzmann free energy functional augmented by our new
  correction accounts for the correlations present in this system when
  compared to molecular dynamics simulations.
\end{abstract}





\newcommand{\rd}{{\operatorname{d}}}
\newcommand{\re}{{\operatorname{e}}}
\newcommand{\rp}{{\operatorname{p}}}
\newcommand{\rB}{{\operatorname{B}}}
\newcommand{\rD}{{\operatorname{D}}}
\newcommand{\rP}{{\operatorname{P}}}
\newcommand{\rDHH}{{\operatorname{DHH}}}
\newcommand{\rDHHC}{{\operatorname{DHHC}}}
\newcommand{\rOCP}{{\operatorname{OCP}}}
\newcommand{\rPB}{{\operatorname{PB}}}

\newcommand{\VECr}{{\boldsymbol{r}}}

\newcommand{\D}{\displaystyle}

\renewcommand{\And}{{{\rm and} }}
\newcommand{\Name}[1]{{{\sc #1}, }}
\newcommand{\Review}[1]{{{\em #1}, }}
\newcommand{\Book}[1]{{{\em #1}, }}
\newcommand{\Vol}[1]{{\bf #1}}
\newcommand{\Year}[1]{{(#1)}}
\newcommand{\Page}[1]{{#1}}


\noindent
The classical one-component plasma (OCP) is an idealized model, in which a
single species of ions moves in a homogeneous neutralizing background of
opposite charge and interacts only via a repulsive Coulomb potential
\cite{Sa58,Abe59,Ba80,Mi87}.  Apart from its applications in plasma physics
\cite{Ro87,Ze92} it is also commonly used in soft matter physics as one of the
simplest possible approaches for modeling correlations when studying
polyelectrolytes, charged planes \cite{St90,Ba99a} or charged colloids
\cite{Al84,Ke86,Pe90,Gr91,Ba99b}.  The general idea is the following: Compute
the OCP free energy as a function of bulk density $n_\rB$ and use this
expression in the spirit of a {\em local density approximation\/} (LDA) as a
correlation correction for the inhomogeneous system (i.e., $n_\rB\rightarrow
n(\VECr)$). The total excess free energy is the volume integral over the free
energy density and thus becomes a functional of $n(\VECr)$.  Many alternative
and more sophisticated methods based on integral equations \cite{Sp73} have
been developed for treating this correlation problem.  Even though they offer
results which are in good agreement with Monte-Carlo simulations, they do not
provide any intuitive insight into the physics governing ionic solutions.

There is, however, a fundamental problem with the local density approaches:
The OCP free energy is not a convex function of density \cite{LiNa75}. This
implies that it cannot be used in a thermodynamically stable way within LDA,
since the system can lower its total free energy by developing local
inhomogeneities and increasing its density in one region at the expense of
another (disregarding any surface effects) \cite{Cal85}. Once started, this
continues as a runaway process and the overall system collapses to a point.
This feature is already seen on the level of the Debye-H{\"u}ckel plus hole
(DHH) approximation \cite{No84}, which is an extension of the original
Debye-H{\"u}ckel (DH) theory \cite{DH23} for the special case of the OCP, and
the instability it gives rise to has been termed ``structuring catastrophe''
in this context \cite{Pe90,Gr91}.  The proper way for avoiding this difficulty
thus requires modifications of the one-component plasma model {\em itself}.
The new theory, referred to as the Debye-H{\"u}ckel-Hole-Cavity (DHHC)
approach, remains simple and can be used within LDA to account for correlation
effects present in more complex ionic solutions, as will be shown in an
example at the end of the paper, where we compare its predictions to
simulational results of a model system.

\vspace*{0.5ex}

Since the necessary changes to DHH will turn out to be surprisingly
tiny, it is worthwhile to briefly recall the way in which DHH theory
arrives at a free energy for the OCP.  For definiteness, we assume a
system of $N$ identical point-particles of valence $v$ and (positive)
unit charge $q$ inside a volume $V$ with a uniform neutralizing
background of density $v n_\rB$ and dielectric constant $\varepsilon$.
According to the DH approach, the potential $\phi$ created by a
central ion ({\em i.e.}, fixed at the origin) and all its surrounding
ions results from solving the spherically symmetric Poisson equation
\begin{equation}
  \label{eq:sph_Poiss}
  \nabla^2 \phi(r) \; = \;
  \phi''(r) + \frac{2}{r}\,\phi'(r) \; = \;
  -\frac{4\pi}{\varepsilon } \rho(r)
\end{equation}                                
under the requirement that the charge density is $\rho(\VECr)=q
v\delta(\VECr)$ at the central ion and that the rest of the mobile
ions rearrange themselves in the uniform background in accordance with
the Boltzmann distribution $\rho(r)= vqn_\rB \exp{[-\beta v q
\phi(r)]} - vqn_\rB$. Combining this with eq.~(\ref{eq:sph_Poiss})
yields the nonlinear Poisson-Boltzmann (PB) equation, while
linearization of the exponential function in the mobile ion density
gives $\rho(r)=-\varepsilon\kappa^2 \phi(r)/4\pi$ together with the
famous Debye-H{\"u}ckel solution for the potential, $\phi(r)\thicksim
\re^{-\kappa r}/r$, illustrating the rearrangement of the other ions
around the central one in order to screen the Coulomb
interaction. Here, $\kappa\equiv \sqrt{4\pi\ell n_\rB}$ is the inverse
screening length, $\ell=\ell_\rB v^2$, with $\ell_\rB=\beta
q^2/\varepsilon$ being the Bjerrum length, and $\beta = 1/k_\rB T$.

The problem with the DH theory is that the condition for linearization
is obviously not satisfied for small $r$, where the potential is large
--- indeed, the particle density becomes negative and finally diverges
at the origin. This defect was overcome by the DHH theory \cite{No84},
which artificially postulates a correlation hole of radius $h$ around
the central ion where no other ions are allowed. In this case the
charge density is given by
\begin{equation}
  \label{eq:rho_DHH}
  \rho(r) \; = \; \left\{
    \begin{array}{r@{\quad:\quad}l}
      q v\,(\delta(\VECr) - n_\rB) & r \le h \\ 
      -\varepsilon \kappa^2 \phi(r)/4\pi & r > h.
    \end{array}
  \right.
\end{equation}
The solution of the linearized PB equation with the appropriate
boundary conditions (continuity of electric field and potential)
yields the potential for both regions in dependence of $h$, which has
to be fixed on physical grounds: At low temperatures the electrostatic
repulsion dominates and the minimum ion separation essentially becomes
the mean separation, so $h=(4\pi n_\rB/3)^{-1/3}$. At high
temperatures, the hole size can be estimated by balancing Coulombic
and thermal energy, which gives $h=\ell$. A systematic way to
interpolate between these two limits results from excluding particles
from a region where their potential energy is larger than some
threshold.  A natural choice is the thermal energy $k_\rB T$, which
leads to
\begin{equation}
  \label{eq:h_DHH_omega}
  \kappa h = \omega-1
  \qquad\text{with}\qquad
  \omega = (1+3\ell\kappa)^{1/3}.
\end{equation}
Incidentally, this assumption also gives a continuous charge density
across the hole boundary.
 
Once the potential at the position of the central ion is known, the
electrostatic contribution to the Helmholtz free energy density can be
obtained by the Debye charging process \cite{DH23}, as was done
previously by Penfold {\em et al.\/} \cite{Pe90}:
\begin{equation}
  \label{eq:f_DHH}
  \frac{\beta f_\rDHH}{n_\rB} \; = \; \frac{1}{4}\left[ 1-\omega^2+\frac{2\pi}{3\sqrt{3}}
    + \ln\left(\frac{\omega^2+\omega+1}{3}\right)
    -\frac{2}{\sqrt{3}}\arctan\left(\frac{2\omega+1}{\sqrt{3}}\right)
  \right].
\end{equation}

\begin{figure}[t]
  \vspace*{-0.5cm}
  \begin{center} \input{fig1} \end{center} \caption{Free
  energy density of the DHH (dashed, eq.~(\ref{eq:f_DHH})) and DHHC
  (solid, eq.~(\ref{eq:f_DHHC})) theory as a function of density for
  Bjerrum length $\ell_\rB=7.14$\AA\ and monovalent/divalent ions
  (left/right). The arrows mark the points at which the DHH free
  energy density changes from convex to concave. A particle volume of
  $V_\rp=(5\mbox{\AA})^3$ was assumed.}\label{fig:f_DHH_DHHC}
\end{figure}
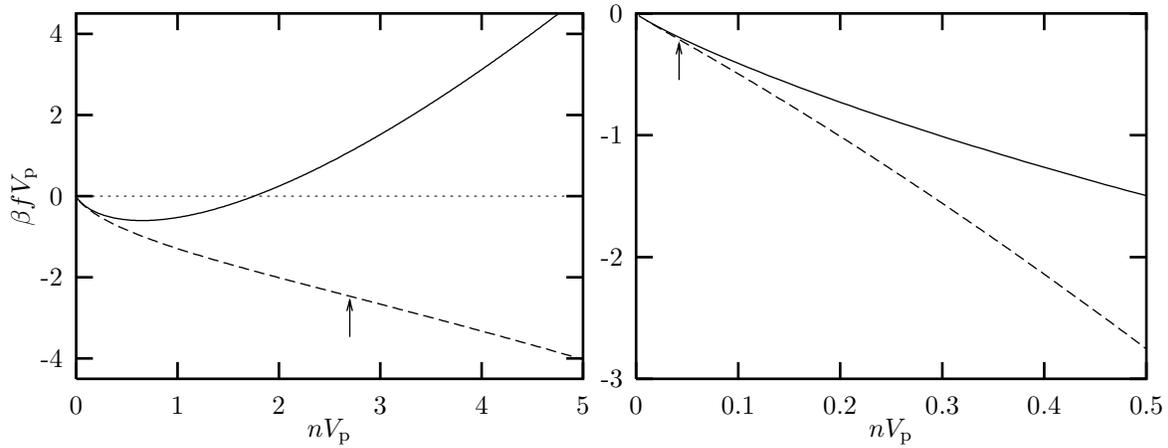

The presented simple DHH analysis of the one-component plasma theory
offers considerable insight into ionic systems and is in good
agreement with Monte-Carlo simulations \cite{Br66} when fluctuations
on the charge density are not relevant \cite{Ta99}.  In principle one
can attempt to include such fluctuations by applying the bulk
density-functional theory in a {\em local\/} way, {\em i.e.},
$n_\rB\rightarrow n(\VECr)$.  The basic idea is to obtain the density
distribution via functional minimization of the Helmholtz free energy
\begin{equation}
  \label{eq:F_OCP}
  \beta F_\rOCP[n(\VECr)] \; = \; \int\!\rd^3r
  \, \Big\{ n(\VECr) \ln\big(n(\VECr) V_\rp\big) 
  \, + \, \beta f_\rDHH[n(\VECr)] \Big\}
\end{equation}
under the constraint of global charge neutrality ($V_\rp$ represents
the volume of a particle.)  Yet, this variational process does not
lead to a well defined density profile, since $f_\rDHH(n)$
asymptotically behaves like $-n^{4/3}$ at high densities and therefore
is not a convex function -- with the implications already mentioned in
the introduction.  At {\em small\/} densities, however, the free
energy density is {\em convex\/} and changes to a concave form only
beyond a critical density $n^\star\approx 7.8618/\ell^3$ (see
fig.~\ref{fig:f_DHH_DHHC}).  Hence, if during the process of actually
computing $n(\VECr)$ such a density is never met, the theory does not
``realize'' its asymptotic instability and gives a finite (yet,
meta-stable) answer. It has in fact been applied to account for
correlations in the case of systems with low ionic strength
\cite{St90,Ba99a,Pe90}).  Assuming the case of aqueous solutions
($\ell_\rB=7.14$\AA) and monovalent ions we find a critical density
$n^\star\approx\;36\,\text{mol/l}$, which clearly is high enough to
prevent a runaway process to set in. However, already for divalent and
trivalent ions we find $n^\star \approx 0.56\,\text{mol/l}$ and
$0.049\,\text{mol/l}$, respectively, which are sufficiently low to be
realized and thus to trigger a collapse. Notice the strong dependence
of $n^\star$ on valence, namely, on the sixth power.

To circumvent the instabilities occurring at high densities
erroneously attributed to the local density approach {\em itself}, a
number of nonlocal free energies have been proposed
\cite{St90,Ba99a,Gr91,Pa99}. In these {\em weighted density
approximations\/} (WDA) the local density is replaced by a spatially
averaged quantity.  The main problem with these methods is that the
choice of the weighting function is somewhat arbitrary.  In most cases
it is obtained by relating the second variation of the free energy
with the direct correlation function.  At this point the WDA requires
prior information about this function, which is not yet available and
thus has to be calculated using different approaches (like, {\em
e.g.}, integral equation theories).  Whatever choice one takes, it is
still {\em (i)\/} quite arbitrary and {\em (ii)\/} leads to a series
of approximations which {\em (iii)\/} instead of clarifying the
physics tend to obscure it.
 
The instabilities present in the local DHH approach can be properly
overcome by recognizing that the failure of this model is due to the
(too strong) requirement of {\em local\/} charge neutrality imposed by
the LDA: A local fluctuation leading to an increase of {\em
particle\/} density implies a corresponding increase in {\em
background} density; therefore the fluctuation is not suppressed by an
increase in repulsive Coulomb interactions but quite on the contrary
favored by its decrease.  A natural solution for that problem is to
decouple the particle density from the background density and to apply
the LDA just to the former one. This, however, leads to nonlinearities
in the solution of the differential equation which spoil the
simplicity of the DH and DHH approximations.  The most simple solution
is to exclude the neutralizing background from a {\em cavity\/} of
radius $a$ placed around the central ion {\em only}, which is already
sufficient to control the unphysical divergence of the particle
density.  Even though it does not accounts for excluded-volume effects
\cite{Ne99,Pe93}, the parameter $a$ can in principle be identified
with the diameter of the particles.  In addition, the exclusion hole
for $a \leq r \leq h$ is retained in order to account for the
electrostatic repulsion between two ions.  Consequently, the charge
density, which for the usual DHH theory is given by
eq.~(\ref{eq:rho_DHH}), has now {\em three\/} regions:
\begin{equation}
  \label{eq:rho_DHHC}
  \rho(r) \; = \; \left\{
    \begin{array}{r@{\quad:\quad}l}
      q v\delta(\VECr) & r < a \\
      -q v n_\rB & a \le r \le h \\[0.3ex]
      -\varepsilon \kappa^2 \phi(r)/4\pi & r > h.
    \end{array}
  \right.
\end{equation}
The solution of the linearized PB equation with appropriated boundary
conditions gives the potential in those regions:
\begin{equation}
  \label{eq:phi_DHHC}
  \hspace*{-2ex}
  \psi(r) = \frac{ve_0}{4\pi\varepsilon r} \times \left\{
    \begin{array}{l@{\quad:\quad}l}
      1-\D\frac{r}{2\ell}\left[(\kappa h)^2-(\kappa a)^2\right] 
      -\kappa r C_h & 0 \le r < a \\[2ex]
      1 -\D\frac{r}{2\ell}\left[(\kappa h)^2-(\kappa r)^2\right]
      -\D\frac{1}{3\ell\kappa}\left[(\kappa r)^3-(\kappa a)^3\right]
      - \kappa r C_h & a \le r < h \\[3ex] 
      C_h \, \re^{-\kappa(r-h)} & h \le r < \infty,
    \end{array}
  \right.
\end{equation}
with the abbreviation
\begin{equation}
  \label{eq:C_h}
  C_h \; = \; \frac{1}{1+\kappa h}\,\left(1 - 
    \frac{(\kappa h )^3-(\kappa a)^3}{3\ell\kappa}\right).
\end{equation}
In order to obtain the old theory in the limit $a \rightarrow 0$ we
choose the hole size $h$ to yield the same screening ({\em i.e.}, the
same amount of charge within $h$) as the DHH theory, which results in
\begin{equation}
\label{eq:h_DHHC}
\kappa h = \big[(\omega-1)^3+(\kappa a)^3\big]^{1/3}.
\end{equation}
This expression has four important physical limits: zero/infinite
temperature and low/high density.  At low temperature the exclusion
hole has maximum size and, like in the DHH case, behaves as $h=(3/4\pi
n_\rB+a^3)^{1/3}$.  As the temperature is increased, the hole size
shrinks, but contrary to DHH theory it does not vanishes and $h
\rightarrow a$ as $T \rightarrow \infty$.  At small densities,
entropic effects compete with the Coulombic repulsion and $h=\ell+a$;
for high densities, the exclusion hole decreases but is again limited
below and $h \rightarrow a$.  Using this prescription for $h$, the
Helmholtz free energy can be obtained by Debye-charging the fluid:
\begin{equation}
  \label{eq:f_DHHC}
  \hspace{-5ex}
  \frac{\beta f_\rDHHC}{n_\rB} \; = \; \frac{(\kappa a)^2}{4} - 
  \int_{1}^{\omega} \! \rd\bar{\omega} \, \Big\{
  \frac{\bar{\omega}^2}{2(\bar{\omega}^3-1)}\Omega(\bar{\omega})^{2/3}
  +\frac{\bar{\omega}^3}{(1+\Omega(\bar{\omega})^{1/3})
    (\bar{\omega}^2+\bar{\omega}+1)} \Big\}
\end{equation}
with the abbreviation
\begin{equation}
  \label{eq:Omega}
  \Omega(\bar{\omega}) = (\bar{\omega}-1)^3
  +\frac{(\kappa a)^3}{3 \ell\kappa}\,(\bar{\omega}^3-1)
\end{equation}
and where $\omega$ is the same as in eq.~(\ref{eq:h_DHH_omega}).  The
integral can be solved numerically for given values of $\ell_\rB$, $v$
and $a$.  As in the DHH approach fluctuations are taken into account
by allowing the density to become local; thus, $n(\VECr)$ is obtained
by minimizing the free energy from eq.~(\ref{eq:F_OCP}) with $f_\rDHH$
replaced by $f_\rDHHC$ as given by eq.~(\ref{eq:f_DHHC}).  But
differently from the DHH theory, the Debye-H{\"u}ckel-Hole-Cavity free
energy is a convex function of density and thus applicable within a
local density approximation.  This situation is depicted in
fig.~\ref{fig:f_DHH_DHHC}, where we plotted the previous expression of
the free energy of the DHH theory together with the improved
expression of the DHHC approach. Recall that the DHH free energy has a
point of inflection at a critical density $n^\star \approx
7.8618/\ell^3$, which makes it unstable at high densities --
particularly for multivalent ionic correlations, as is demonstrated in
the right part of fig.~\ref{fig:f_DHH_DHHC}.

\vspace{0.5ex}

As an example, we apply this free energy as a correlation correction
in the theoretical description of the screening of a charged rod,
which is a simple model of biologically relevant stiff
polyelectrolytes like DNA, actin filaments or microtubules. Much of
the thermodynamic behavior of these molecules is determined by the
distribution of the counterions around the polyion.  As a model system
we take a rod of radius $r_0$ and line charge density
$\lambda=0.959\,q/r_0$ embedded in a cell of outer radius
$R=123.8\,r_0$ and the complementary values $\ell_\rB/r_0=3$, $v=1$
and $\ell_\rB/r_0=1$, $v=3$ have been investigated, which on the plain
PB level both give a fraction of condensed counterion (in the Manning
sense) of roughly 65\% \cite{Manning,De99}.  This system is thus
strongly charged and one expects ionic correlations to become
relevant. Indeed, the comparison between the distributions obtained by
simulation \cite{De99} and the ones from PB theory shows that the
mean-field approach fails in the limit of high ionic strength.  In
reality the ions do not just interact with the average electrostatic
field but if an ion is present in a position $\VECr$, it tends to push
away other ions from that point. This effect becomes important at high
densities, low temperatures and for multivalent ions.

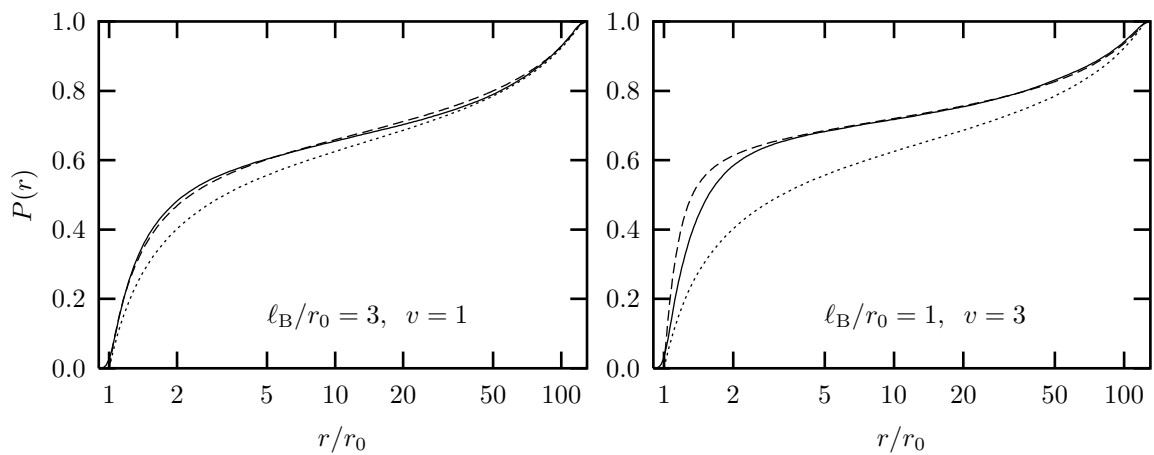
\begin{figure}[t]
  \vspace*{-0.5cm}
  \begin{center} \input{fig2} \end{center}
  \caption{Counterion distribution function $P(r)$ from eq.~(\ref{eq:P(r)}) for 
    two cylindrical cell models with $R/r_0=123.8$, $\lambda=0.959\,q/r_0$ and
    the values for Bjerrum length and valence as indicated in the plots. The
    solid line is the result of a molecular dynamics simulation while the
    dotted line is the prediction from Poisson-Boltzmann theory. The increased
    counterion condensation visible in the simulation is accurately captured
    by the extended PB theory (dashed line) using the correction from
    eq.~(\ref{eq:f_DHHC}).}\label{fig:DHHC_example}
\end{figure}

As discussed above, a simple way to improve PB theory is to extend the
density functional to include a term of the form (\ref{eq:f_DHHC})
which accounts for the correlations.  The configurational free energy
for the screened macroion solution can be partitioned into two terms:
\begin{equation}
  \label{eq:F_P}
  F_\rP[n(\VECr)] \; = \; F_\rPB[n(\VECr)] +
  \int_{}^{} \! \rd^3r \; f_\rDHHC[n(\VECr)].
\end{equation}
The first part
\begin{equation}
  \label{eq:F_PB}
  F_\rPB[n(\VECr)] \; = \; \int \! \rd^3r \, \Big \{
  k_\rB T \, n(\VECr) \ln\big(n(\VECr) V_\rp\big)
  +\frac{1}{2} \, q v \, n(\VECr)\phi[n(\VECr)]
  \Big\}
\end{equation}
contains the ideal gas contribution of the small ions, the interaction
with the macroion potential and the mean-field interaction between the
counterions.  Minimization of this expression under the constraint of
global charge neutrality gives -- together with the Poisson equation
-- the Poisson-Boltzmann equation. The inter-particle correlations are
now approximately accounted for by adding an excess free energy, which
is the second term in eq.~(\ref{eq:F_P}) --- the DHHC free energy in
local density approximation.  The equilibrium ion distribution
minimizing the functional (\ref{eq:F_P}) is most easily found by means
of a Monte-Carlo solver, as has been proposed elsewhere
\cite{Deserno1999b}. The fraction of ions within a distance $r$,
\begin{equation}
  \label{eq:P(r)} P(r) \; = \; \frac{1}{\lambda}\int_{r_0}^{r} \! \rd
  \bar{r} \; 2\pi\bar{r} \, v q \, n(\bar{r}),
\end{equation}
obtained following this procedure is illustrated in
fig.~\ref{fig:DHHC_example}. Compared to the plain PB result the
simulation shows a stronger condensation of ions in the vicinity of
the rod, an effect which is more pronounced in the trivalent
system. In both cases the increased condensation is reproduced by the
correlation corrected PB functional from eq.~(\ref{eq:F_P}). While in
the case $\ell_\rB/r_0=3$, $v=1$ the theoretical prediction
practically overlaps the simulation, it somewhat overestimates
correlations in the complementary case $\ell_\rB/r_0=1$, $v=3$.  It
must, however, be noted that the ions in the simulation also
interacted via a repulsive Lennard-Jones potential, giving them a
diameter of roughly $r_0$.  The expected reduction of particle density
resulting from the additional hard core is not accounted for in the
presented theory, but could easily be included along the lines of
Refs.~\cite{Bo97}.

\vspace*{0.5ex}

In conclusion, we have shown that the failure of the local density
approximation for the one-component plasma is due to the
asymptotically concave free energy employed by the DHH theory. To
eliminate this problem, we introduced a DHHC approach in which the
uniform background is absent in the immediate vicinity of the central
ion, which leads to a convex, thermodynamically stable free energy.
Moreover, the local density functional theory derived from this
assumption is able to correctly account for the correlations between
small ions in the presence of a strongly charged macroion. This was
demonstrated for the case of a stiff rodlike polyelectrolyte by
comparing the integrated charge density to simulation results of the
same model. A more detailed investigation of the applicability of the
LDA to rodlike polyelectrolytes and to charged colloids is postponed
to future work.
  

\vspace*{-0.2cm}

\begin{acknowledgments}

\vspace*{-0.2cm}
 
This work has been supported by the Brazilian agency CNPq (Conselho
Nacional de Desenvolvimento Cient{\'\i}fico e
Tecnol{\'o}gico). M. C. B. would like to thank K. Kremer for his
hospitality during a stay in Mainz, where most of this work was
completed.

\end{acknowledgments}


\end{document}

%% file: fig1.tex
\setlength{\unitlength}{0.1bp}
\begin{picture}(4320,1728)(0,0)
\special{psfile=fig1 llx=0 lly=0 urx=864 ury=403 rwi=8640}
\put(3321,50){\makebox(0,0){$n V_\rp$}}
\put(4283,150){\makebox(0,0){0.5}}
\put(3898,150){\makebox(0,0){0.4}}
\put(3514,150){\makebox(0,0){0.3}}
\put(3129,150){\makebox(0,0){0.2}}
\put(2745,150){\makebox(0,0){0.1}}
\put(2360,150){\makebox(0,0){0}}
\put(2310,1628){\makebox(0,0)[r]{0}}
\put(2310,1169){\makebox(0,0)[r]{-1}}
\put(2310,709){\makebox(0,0)[r]{-2}}
\put(2310,250){\makebox(0,0)[r]{-3}}
\put(1205,50){\makebox(0,0){$n V_\rp$}}
\put(75,939){%
\special{ps: gsave currentpoint currentpoint translate
270 rotate neg exch neg exch translate}%
\makebox(0,0)[b]{\shortstack{$\beta f V_\rp$}}%
\special{ps: currentpoint grestore moveto}%
}
\put(2160,150){\makebox(0,0){5}}
\put(1778,150){\makebox(0,0){4}}
\put(1396,150){\makebox(0,0){3}}
\put(1014,150){\makebox(0,0){2}}
\put(632,150){\makebox(0,0){1}}
\put(250,150){\makebox(0,0){0}}
\put(200,1551){\makebox(0,0)[r]{4}}
\put(200,1245){\makebox(0,0)[r]{2}}
\put(200,939){\makebox(0,0)[r]{0}}
\put(200,633){\makebox(0,0)[r]{-2}}
\put(200,327){\makebox(0,0)[r]{-4}}
\end{picture}

%% file: fig2.tex
\setlength{\unitlength}{0.1bp}
\begin{picture}(4320,1728)(0,0)
\special{psfile=fig2 llx=0 lly=0 urx=864 ury=403 rwi=8640}
\put(3056,516){\makebox(0,0)[l]{$\ell_\rB/r_0 = 1, \;\; v=3$}}
\put(3346,50){\makebox(0,0){$r/r_0$}}
\put(4184,220){\makebox(0,0){100}}
\put(3923,220){\makebox(0,0){50}}
\put(3578,220){\makebox(0,0){20}}
\put(3317,220){\makebox(0,0){10}}
\put(3056,220){\makebox(0,0){5}}
\put(2711,220){\makebox(0,0){2}}
\put(2450,220){\makebox(0,0){1}}
\put(2360,1628){\makebox(0,0)[r]{1.0}}
\put(2360,1366){\makebox(0,0)[r]{0.8}}
\put(2360,1105){\makebox(0,0)[r]{0.6}}
\put(2360,843){\makebox(0,0)[r]{0.4}}
\put(2360,582){\makebox(0,0)[r]{0.2}}
\put(2360,320){\makebox(0,0)[r]{0.0}}
\put(954,516){\makebox(0,0)[l]{$\ell_\rB/r_0 = 3, \;\; v=1$}}
\put(1240,50){\makebox(0,0){$r/r_0$}}
\put(65,974){%
\special{ps: gsave currentpoint currentpoint translate
270 rotate neg exch neg exch translate}%
\makebox(0,0)[b]{\shortstack{$P(r)$}}%
\special{ps: currentpoint grestore moveto}%
}
\put(2063,220){\makebox(0,0){100}}
\put(1806,220){\makebox(0,0){50}}
\put(1467,220){\makebox(0,0){20}}
\put(1211,220){\makebox(0,0){10}}
\put(954,220){\makebox(0,0){5}}
\put(615,220){\makebox(0,0){2}}
\put(359,220){\makebox(0,0){1}}
\put(270,1628){\makebox(0,0)[r]{1.0}}
\put(270,1366){\makebox(0,0)[r]{0.8}}
\put(270,1105){\makebox(0,0)[r]{0.6}}
\put(270,843){\makebox(0,0)[r]{0.4}}
\put(270,582){\makebox(0,0)[r]{0.2}}
\put(270,320){\makebox(0,0)[r]{0.0}}
\end{picture}

%% file: paper.bbl
\vspace*{-0.5cm}

\begin{thebibliography}{99}

\vspace*{-1.5cm}


\bibitem{Sa58} \Name{Salpeter E. E.}  \Review{Ann. Phys.}  \Vol{5} \Year{1958}
  \Page{183}.
%
\bibitem{Abe59} \Name{Abe R.}  \Review{Prog. Theory Phys.}  \Vol{22}
  \Year{1959} \Page{213}.
%
\bibitem{Ba80} \Name{Baus M. \And Hansen J.-P.}  \Review{Phys. Rep.}  \Vol{59}
  \Year{1980} \Page{1}.
%
\bibitem{Mi87} \Name{Minnhagen P.} \Review{Rev. Mod. Phys.}  \Vol{59}
  \Year{1987} \Page{1001}.
%
\bibitem{Ro87} \Name{Rogers F. J. \And DeWitt H. E.} eds., \Book{Strongly
    Coupled Plasmas} Plenum Press, New York, \Year{1987}.
%
\bibitem{Ze92} \Name{Zerah G., Clerouin J. \And Pollock E. L.}  \Review{Phys.
    Rev. Lett.} \Vol{69} \Year{1992} \Page{446}.
%
\bibitem{St90} \Name{Stevens M. J. \And Robbins M. O.}  \Review{Europhys.
    Lett.} \Vol{12} \Year{1990} \Page{81}.
%
\bibitem{Ba99a} \Name{Diehl A., Barbosa M. C., Tamashiro M. N. \And Levin Y.}
  to appear in {\em Physica A}. {\sf cond-mat/9904114}.
%
\bibitem{Al84} \Name{Alexander S., Chaikin P. M., Grant P., Morales G. J.,
    Pincus P. \And Hone D.} \Review{J. Chem. Phys.} \Vol{80} \Year{1984}
  \Page{5776}.
%
\bibitem{Ke86} \Name{Kremer K., Robbins M. O. \And Grest G. S.}  \Review{Phys.
    Rev. Lett.} \Vol{57} \Year{1986} \Page{2694}.
%
\bibitem{Pe90} \Name{Penfold R., Nordholm S., J{\"o}nsson B. \And Woodward C.
    E.} \Review{J. Chem. Phys.} \Vol{92} \Year{1990} \Page{1915}.
%
\bibitem{Gr91} \Name{Groot R.} \Review{J. Chem. Phys.} \Vol{95} \Year{1990}
  \Page{9191}.
%
\bibitem{Ba99b} \Name{Levin Y., Barbosa M. C. \And Tamashiro M. N.}
  \Review{Europhys. Lett.} \Vol{41} \Year{1998} \Page{123}; \Name{Tamashiro M.
    N., Levin Y. \And Barbosa M. C.} \Review{Physica A} \Vol{258} \Year{1998}
  \Page{341}.
  
\bibitem{Sp73} \Name{Springer J. F., Porkrant M.-A. \And Stevens F. A.}
  \Review{J. Chem. Phys.} \Vol{58} \Year{1973} \Page{4863}; \Name{Ng K.-C.}
  \Review{J. Chem. Phys.} \Vol{61} \Year{1974} \Page{2680}; \Name{Rosenfeld
    Y.} \Review{Phys. Rev. E} \Vol{54} \Year{1996} \Page{2827}.
%
\bibitem{LiNa75} \Name{Lieb E. H. \And Narnhofer H.} \Review{J. Stat. Phys.}
  \Vol{12} \Year{1975} \Page{291}. Erratum: \Review{J. Stat. Phys.} \Vol{14}
  \Year{1976} \Page{465}.
%
\bibitem{Cal85} \Name{Callen H. B.} \Book{Thermodynamics and an Introduction
    to Thermostatistics} 2nd ed., Wiley, New York, \Year{1985}, chapter 8.1.
%
\bibitem{No84} \Name{Nordholm S.} \Review{Chem. Phys. Lett.} \Vol{105}
  \Year{1984} \Page{302}.
%
\bibitem{DH23} \Name{Debye P. \And H{\"u}ckel E.} \Review{Phys. Z.} \Vol{24}
  \Year{1923} \Page{185}; \Page{305}; \Name{McQuarrie D. A.} \Book{Statistical
    Mechanics} Harper-Collins, New York, \Year{1976}, chapter 15.
%
\bibitem{Br66} \Name{Brush S. G., Sahlin H. L. \And Teller E.}  \Review{J.
    Chem. Phys.} \Vol{45} \Year{1966} \Page{2102}.
%
\bibitem{Ta99} \Name{Tamashiro M. N., Levin Y. \And Barbosa M. C.}
  \Review{Physica A} \Vol{268} \Year{1999} \Page{24}.
%
\bibitem{Pa99} \Name{Patra C. N. \And Yethiraj A.} \Review{J. Phys. Chem.}
  \Vol{103} \Year{1999} \Page{6080}.
%
\bibitem{Ne99} \Name{Netz R. R. \And Orland H.} {\sf cond-mat/9902220}.
%
\bibitem{Pe93} \Name{Penfold R., J{\"o}nsson B. \And Nordholm S.}  \Review{J.
    Chem. Phys.} \Vol{99} \Year{1993} \Page{497}.
%
\bibitem{Manning} \Name{Manning G. S.} \Review{J. Chem. Phys.} \Vol{51}
  \Year{1969} \Page{924}; \Page{934}; \Page{3249}.
%
\bibitem{De99} \Name{Deserno M., Holm C. \And May S.}  {\sf cond-mat/9906277},
  to appear in {\em Macromolecules}.
%
\bibitem{Deserno1999b} \Name{Deserno M.} {\sf cond-mat/9910282}, submitted to
  {\em Physica A}.
%
\bibitem{Bo97} \Name{Borukhov I., Andelman D. \And Orland H.}  \Review{Phys.
    Rev. Lett.} \Vol{79} \Year{1997} \Page{435}. \Name{Lue L., Zoeller N. \And
    Blankschtein D.} \Review{Langmuir} \Vol{15} \Year{1999} \Page{3726}.
%
\end{thebibliography}
